\documentclass[conference]{IEEEtran}
\usepackage{cite}
\usepackage{amsmath,amssymb,amsfonts}
\usepackage{algorithmic}
\usepackage{graphicx}
\usepackage{textcomp}
\usepackage{pgfplotstable}
\usepackage{booktabs} 
\usepackage{tabularx}
\usepackage{etoolbox}
\usepackage{multirow}
\usepackage{float}

\usepackage{hyperref}

\usepackage[dvipsnames]{xcolor}

\newif\ifshowtodos
\showtodosfalse 
\makeatletter 
\newcommand{\@todo}[3]{%
  \ifshowtodos
    \textcolor{#2}{ToDo #1: #3}%
  \fi}

\newcommand{\todomk}[1]{\@todo{MK}{BrickRed}{#1}}
\newcommand{\todomc}[1]{\@todo{MC}{Sepia}{#1}}
\newcommand{\todorg}[1]{\@todo{RG}{Violet}{#1}}
\newcommand{\todokw}[1]{\@todo{KW}{Blue}{#1}}
\newcommand{\todowa}[1]{\@todo{WA}{MidnightBlue}{#1}}
\newcommand{\todojl}[1]{\@todo{JL}{ForestGreen}{#1}}
\newcommand{\todost}[1]{\@todo{ST}{Mahogany}{#1}}
\makeatother   

\usepackage{fancyhdr}
\usepackage{listings}
\pgfplotsset{compat=1.18}
\def\BibTeX{{\rm B\kern-.05em{\sc i\kern-.025em b}\kern-.08em
    T\kern-.1667em\lower.7ex\hbox{E}\kern-.125emX}}

\showtodostrue
    
\begin{document}

\title{Multimodal Speaker Identification in Classroom Environments\\}

\author{
\IEEEauthorblockN{
Michael Leon Chrzan\textsuperscript{1}, 
Meghavarshini Krishnaswamy\textsuperscript{1}, 
Robert Gibboni\textsuperscript{2},\\ 
Katie Wetstone\textsuperscript{2},
Wei Ai\textsuperscript{3}, and 
Jing Liu\textsuperscript{1}
}
\IEEEauthorblockA{
\textsuperscript{1}\textit{Center for Educational Data Science and Innovation, University of Maryland}\\
\textsuperscript{2}\textit{DrivenData}\\
\textsuperscript{3}\textit{College of Information and the Institute for Advanced Computer Studies, University of Maryland} 
}
\IEEEauthorblockA{
\textit{Corresponding Authors:}\\
mlchrzan@umd.edu, jliu28@umd.edu
}
}

\maketitle
\begin{abstract}
Automated analysis of K-12 classroom dynamics faces challenges due to background noise and variable child speech, often confounding acoustic-only models. This study evaluates a multimodal speaker identification framework anchoring acoustic embeddings with LLM-derived semantic context. Using a subset of the EDSI dataset (8 math classrooms, $N=2,801 \text{ utterances}$), we found an acoustic baseline (ECAPA-TDNN) achieved only 39.0\% accuracy. By integrating transcript-based "contextual anchoring" into a gradient boosting classifier, our multimodal approach raised student identification to 50.3\%. Performance also improved for utterances over 5 seconds, reaching 76.9\% accuracy (vs. 64.9\% baseline) with a 90.9\% Top-3 accuracy. Additionally, the model distinguished teacher vs. student roles with 99.3\% accuracy. This approach advances the feasibility of automated feedback systems capable of considering individual student participation, a crucial step for supporting equitable instruction at scale.
\end{abstract}

\begin{IEEEkeywords}
speaker identification, multimodal, K-12 classrooms
\end{IEEEkeywords}

\section{Introduction}
The automated analysis of classroom discourse has emerged as a pivotal frontier in educational technology, promising to provide scalable, objective feedback to teachers regarding instructional quality, student engagement, and equity of participation \cite{xu_promises_2024, demszky_automated_2025, demszky_measuring_2021, dorottya_demszky_m-powering_2023, demszky_can_2024}. Central to this analysis is the task of speaker identification (SID) which seeks to answer the fundamental questions of ``who spoke when'' and ``who said what'' \cite{kanda_transcribe--diarize_2022}. While traditional systems have relied heavily on acoustic biometrics, recent advancements in deep learning and Large Language Models (LLMs) suggest that a paradigm shift towards multimodal frameworks is useful for addressing the unique and adversarial acoustic conditions of the K-12 classroom.

The classroom environment presents a ``perfect storm'' of challenges for conventional audio processing: high levels of non-stationary background noise, significant reverberation, and the distinctive, highly variable spectral characteristics of children's speech. Purely acoustic approaches frequently falter in these settings, struggling to distinguish between target speakers and the pervasive ``babble noise'' of peer discussions \cite{southwell_challenges_2022, xu_promises_2024, tabatabaee_ft-boosted_2025, ma_challenge_2021}. 

While acoustic models provide physical evidence of identity, we propose an additional LLM Speaker Inference component, which introduces evidence derived from the semantic content of the speech. The literature identifies this as a burgeoning field of ``LLM-adaptive diarization'', where language models are used to correct, refine, and attribute speaker turns based on context \cite{thebaud_enhancing_2025, meizlish_evaluating_2025, efstathiadis_llm-based_2025}. This research focuses precisely on how pre-existing transcripts can serve as input for LLMs to derive advanced stylistic features, which are then integrated with high-fidelity acoustic embeddings. 

We work to answer the following question: what is the comparative performance gain achieved by supplementing audio embeddings with context-rich textual features (teacher cues and diarization labels) versus a purely acoustic feature set, particularly concerning the identification of non-teacher speakers? To that end, our study demonstrates a process for incorporating information from transcripts of classroom speech with acoustic models to improve speaker ID performance.

\begin{figure}[htbp]
    \centering
    \label{fig:justification}
    \caption{Speaker Identification Uses}
    \includegraphics[width=0.45\textwidth]{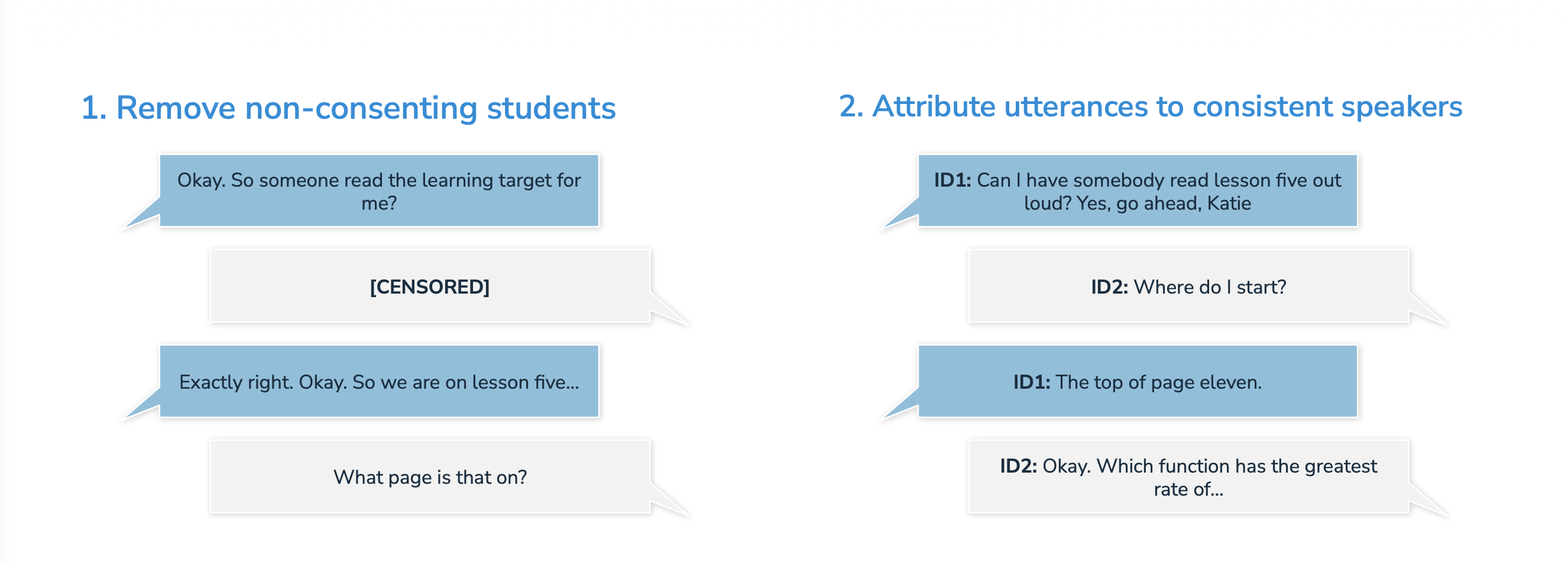}
    {\raggedright\footnotesize Speaker identification in this study served two main purposes, as displayed in this figure: (1) To adhere to legal and ethical responsibilities towards students who did not consent to be part of the benchmark dataset we needed a way to identify and anonymize their speech data and (2) Track individual students throughout a session to enable detailed research about classroom speaking patterns and interactions. \par}
\end{figure}

\section{Related Work}
Speaker identity is an inherent multimodal construct defined by two complementary biometric elements: the acoustic signature and the linguistic signature \cite{loweimi_usefulness_2024}. The acoustic signature relates to the physical and physiological characteristics of the vocal tract, determining pitch, timbre, and prosody. The linguistic signature, conversely, encapsulates the speaker's acquired, high-level behavioral habits, including syntactic complexity, preferred vocabulary, discourse markers, and pragmatic strategies. These linguistic choices, often studied under the umbrella of computational sociolinguistics, are stable markers unique to an individual’s education, dialect, and social context.

Speaker Identification is a foundational task in speech processing, essential for biometric authentication, forensic analysis, and personalized conversational AI systems \cite{efstathiadis_llm-based_2025, thebaud_enhancing_2025}. Historically dominated by acoustic signal processing, the field is undergoing a paradigm shift driven by the integration of Large Language Models (LLMs), enabling the system to leverage linguistic context alongside vocal characteristics. 

A speaker’s identity is encoded not only in \textit{what} they say (lexical content) but also in \textit{how} they structure the conversation and manage social interaction. LLMs are uniquely equipped to process these complex, context-sensitive linguistic phenomena. 
LLMs can infer identity through ``contextual anchoring''. If a teacher nominates a student by name in a given utterance (``What do you think, Jason?''), the LLM can identify the addressee of the utterance, and predict that the  subsequent speaker is highly likely to be ``Jason''. 
Similarly, if the transcript contains ``Mr. Smith, can you help me?'', the addressee is Mr. Smith (Teacher), the speaker is highly likely to be a student and the next speaker is likely to be Mr. Smith.
In both these examples, the named addressee can also be eliminated from the list of potential speakers.
This transforms the problem from unsupervised clustering to semi-supervised tracking, using Named Entity Recognition (NER) to create ``anchors'' that constrain the acoustic clustering algorithm \cite{nguyen_identifying_2024}. 
Thus, LLMs can move beyond simple semantic tagging to identify the functional components of speech acts.


Recent work has begun to explore multimodal approaches to address the acoustic challenges of the classroom. For example, Perez et al. (TeachFX) demonstrated that integrating LLM-based re-scoring with acoustic clustering significantly improved the distinction between teacher and student speech. However, while their approach focuses on binary role classification (Teacher vs. Student), our work extends this multimodal framework to the task of granular speaker identification. We leverage 'contextual anchoring' within the transcript not just to separate roles, but to distinguish individual students from one another—a significantly more complex task given the acoustic similarity among peer speakers \cite{perez_multimodal_2025}.

In addition, previous work has shown the potential of automated feedback to be transformative for classroom practice. That potential lies in its ability to offer a scalable, efficient, and objective alternative to traditional, resource-intensive instructional measurement. Historically, identifying complex instructional features or discourse strategies required time-consuming and sometimes inconsistent manual analysis by expert raters \cite{demszky_automated_2025, meizlish_evaluating_2025}. However, the growing literature suggests that Natural Language Processing (NLP) techniques provide a transformative approach for instructional measurement. These tools can automate the analysis of classroom transcripts, providing private, on-demand feedback to teachers, which has been shown in some cases to lead to positive impacts on educators’ instruction quality and student outcomes \cite{demszky_automated_2025, demszky_can_2024, dorottya_demszky_can_2021, demszky_measuring_2021, malamut_james_facilitating_nodate}. This efficiency allows for the development of feedback systems that are cost-efficient and can be delivered quickly and frequently. Crucial to the current and future effectiveness and expansiveness of that feedback is the ability to both accurately identify the teacher's speech from the students' and also individual students from one another.

\begin{table*}[t]
    \centering
    \caption{Summary of Teacher-Student Talk Time and Classroom Metrics}
    \label{tab:descriptives}
    \resizebox{\textwidth}{!}{
        \begin{filecontents*}{table1_video_descriptives.csv}
Teacher,Class Size,Session Duration (s),Total Utterances,Avg Teacher Talk Time (s),Avg Student Talk Time (s),Audio Consent Pct,Video Consent Pct
Natalia,35,4296.2,417,{14.33 (19.07)},{1.69 (2.32)},54.3,42.9
Anastasia,28,3625.9,296,{18.8 (19.63)},{3.11 (6.63)},100.0,96.4
Maria,18,4044.6,623,{6.68 (9.29)},{1.19 (1.52)},50.0,38.9
Kathy,32,3826.0,467,{11.55 (15.15)},{1.4 (1.56)},100.0,100.0
Dawn,17,2936.0,686,{4.93 (6.68)},{1.44 (1.84)},88.2,58.8
Mackenzie,15,3995.4,733,{4.88 (7.23)},{2.52 (3.61)},100.0,100.0
Candice,35,2536.9,368,{9.01 (12.04)},{0.75 (0.86)},62.9,54.3
Latoya,35,3149.0,409,{10.8 (16.33)},{1.73 (1.83)},80.0,71.4
Summary,215,28410.0,3999,10.12,1.73,79.4,70.3
\end{filecontents*}
\pgfplotstabletypeset[
            col sep=comma,
            string type,
            text indicator=", 
            every head row/.style={before row=\toprule, after row=\midrule},
            every last row/.style={
                before row=\addlinespace[0.5ex]\midrule, 
                after row=\bottomrule
            }
        ]{table1_video_descriptives.csv}
    }
    \par \vspace{2ex}
    {\raggedright\footnotesize \textit{Note.} s = seconds; Average columns contain Standard Deviations in parenthesis.\par}
\end{table*}

\section{Data}

This study utilizes data from the Center for Educational Data Science and Innovation (EDSI) Dataset, a comprehensive educational repository developed at the University of Maryland–College Park \cite{noauthor_benchmark_nodate}. The EDSI dataset was constructed to advance the application of artificial intelligence (AI) in educational research and practice, specifically within the context of mathematics classrooms. It integrates multimodal data sources to provide a holistic view of classroom dynamics, distinguishing itself through a rich combination of raw observational data and structured metadata.

The core components of the dataset include high-fidelity audio and video recordings of classroom interactions, accompanied by high-quality transcriptions. These observational data are linked with extensive metadata, including:
\begin{itemize}
    \item Demographics: detailed profiles of student and teacher characteristics.
    \item Classroom Artifacts: contextual materials such as seating charts.
    \item Metrics: student achievement data and psychometric survey responses from both students and teachers.
\end{itemize}

Data collection for the EDSI dataset employs a multi-microphone setup to capture the complex auditory environment of active classrooms. The audio is recorded using Swivl robotic platforms equipped with five labeled microphones to track and record audio from different sources simultaneously \cite{noauthor_swivl_nodate}. For transcription purposes, these individual audio streams are combined into a single mixed-audio file. This composite file is then processed by TranscribeMe to generate highly accurate professional-grade transcripts of classroom discourse which included the transcribed speech, start and end times of utterances, and diarization labels \cite{noauthor_transcribeme_nodate}.

For the present study, the analytical sample consists of recordings from eight different mathematics classrooms ($8$) selected from the broader EDSI corpus, as detailed in Table~\ref{tab:descriptives}. These classes were selected for their variation in student participation as well as the different mixture of background noise based on instructional choices (e.g. some teachers had a higher propensity for group work than others). For each of these eight classes, we utilized the mixed-audio file generated during the collection phase along with its corresponding transcript for model building. For establishing ground truth speaker identification, we use these files along with their corresponding video files and administrative data (such as seating charts, class rosters, etc). While the raw dataset contained $3,999$ utterances, the final labeled dataset for training was filtered to $2,801$ utterances following the adjudication process described in our section~\ref{subsec:speakers}.

\section{Methods}
\label{sec:Methods}

We developed an automated speaker identification system to assign individual speaker identities to each utterance in classroom mathematics instruction videos. The system combines speaker embedding models with gradient boosting classifiers to distinguish between teachers and students, and among individual students within a classroom session.

\subsection{Speaker Annotation}\label{AA}
\label{subsec:speakers}

Two trained annotators (including the first author) conducted speaker identification for each utterance in the classroom transcripts. Annotators worked with multiple sources of contextual information: the mixed-audio track, the classroom video recording, the seating chart, and the class roster. These materials were used in combination to maximize the accuracy of speaker attribution.

For each utterance in the transcript, annotators entered responses into an amended transcript file containing additional columns designed for this task. The required codes were as follows:

\begin{itemize}
    \item \textbf{transcript\_file\_name}: The name of the transcript file being annotated.
    \item \textbf{turn\_idx}: The zero-indexed row number of the utterance in the raw transcript file, ensuring alignment with the transcript when read programmatically (e.g., in pandas).
    \item \textbf{identified (0/1)}: A binary indicator of whether the speaker could be matched to the teacher or a consenting student in the class. 
    \item \textbf{first\_name / last\_name}: If identified, the speaker’s first and last name, exactly matching the entries in the administrative data (including capitalization and special characters). If the speaker was not identified, these fields were left blank.
    \item \textbf{unidentified\_type}: If the speaker could not be matched, annotators classified the voice into one of the following categories:
    \begin{itemize}
        \item child: a distinctly child voice not matched to a known student.
        \item adult: a distinctly adult voice not matched to the teacher.
        \item multiple\_students: overlapping speech from 2–3 students collapsed into a single utterance.
        \item whole\_class: more than 2–3 students speaking simultaneously, collapsed into one utterance.
        \item other: voices not representing classroom participants (e.g., loudspeaker announcements, videos).
        \item impossible: turns that could not be attributed to a single speaker (e.g., [crosstalk], [inaudible], or extremely short utterances).
    \end{itemize}
    \item \textbf{notes}: A free-text field for annotators to record idiosyncrasies, uncertainties, or contextual observations during annotation.
\end{itemize}

This structured coding scheme ensured consistency across annotators and facilitated downstream quantitative analysis.

After completing the first round of annotation, the two annotators met to review discrepancies, particularly in classrooms with low inter-rater reliability (IRR) as measured by Cohen’s kappa \cite{cohen_coefficient_1960}. These meetings focused on clarifying ambiguous codes—notably the distinction between \textit{multiple\_students} and \textit{whole\_class}—and resolving systematic sources of disagreement. Following consensus-building discussions, annotators re-annotated the affected lessons to improve reliability and ensure methodological rigor. 

Challenges in the speaker annotation process also arose from incomplete video consent in some classes, which meant that only a subset of students were visible in the recordings (up to over half of the students in some of the courses, see Table~\ref{tab:descriptives}). As a result, annotators had to rely on audio cues, seating chart positions, and contextual inference rather than direct visual confirmation for some utterances in those courses, particularly when non-consenting students spoke more often in those sessions. 

Altogether, the annotators agreed on \textbf{90.92\%} of their annotations for the 3999 utterances and - despite these challenges - were able to confidently identify \textbf{66.7\%} of all utterances with a specific speaker name ($2668$ utterances). Overall and individual session agreement levels can be found in Table~\ref{tab:kappas}, where we reached almost perfect agreement on 6 of the sessions and substantial agreement on 2 sessions, leading to almost perfect agreement overall \cite{landis_measurement_1977}.

To finalize the dataset, we conducted a manual review of the remaining disagreements. This review revealed that the discrepancies were primarily between the generic 'child' label and a specific student name. We found that the primary annotator provided more specific identifications in these instances compared to the second annotator. To reduce label noise and maximize the quantity of labeled data available for training, we decided to adopt the primary annotator's specific labels in these conflicting cases. Consequently, after resolving these disagreements and filtering out unidentifiable turns (such as \textit{whole\_class} or \textit{impossible}) the final ground truth dataset consisted of $N = 2,801$ utterances.

\begin{table}[t]
    \centering
    \caption{Cohen's Kappa by Teacher}
    \label{tab:kappas}
    \begin{filecontents*}{cohens_kappa_by_teacher.csv}
Teacher,Cohens Kappa
Mackenzie,"1.00 [1.00, 1.00]"
Anastasia,"0.96 [0.94, 0.99]"
Maria,"0.95 [0.93, 0.97]"
Candice,"0.91 [0.88, 0.95]"
Latoya,"0.84 [0.80, 0.88]"
Natalia,"0.84 [0.79, 0.88]"
Kathy,"0.79 [0.74, 0.83]"
Dawn,"0.74 [0.70, 0.78]"
Overall,"0.90 [0.89, 0.91]"
\end{filecontents*}
\pgfplotstabletypeset[
        col sep=comma,
        string type,
        text indicator=", 
        every head row/.style={
            before row=\hline, 
            after row=\hline   
        },
        every last row/.style={
            after row=\hline   
        },
        every row no 8/.style={
            before row=\hline
        },
        columns/{Cohen's Kappa}/.style={
            column name={Cohen's Kappa(95\% CI)},
            string type,
            postproc cell content/.append code={
                \pgfkeysalso{@cell content=\expandafter\detokenize\expandafter{\pgfplotsretval}}
            }
        }
    ]{cohens_kappa_by_teacher.csv}
    \par \vspace{2ex}
    {\raggedright\footnotesize \textit{Note.} 95\% Confidence Interval in parens.\par}
\end{table}

\subsection{Feature Engineering}

Given the high inter-rater reliability, we determined that the data was suitable ground truth labels for model training and evaluation and moved forward with model building using features generated from both the text transcripts and audio data.

We first focused on generating speaker embeddings for identifying speakers. Each classroom session included voice enrollment recordings for all consented participants (teacher and students). These recordings, collected at the beginning of the school year, consisted of 30-60 seconds of speech per speaker reading standardized text. Voice enrollments served as reference exemplars for computing speaker embeddings. Using this audio, we segmented each speaker's recording into 3-second overlapping windows (with 0-second overlap) and computed embeddings for each window, creating multiple reference embeddings per enrolled speaker.

We extracted speaker embeddings using the SpeechBrain ECAPA-TDNN model (spkrec-ecapa-voxceleb), a state-of-the-art speaker verification system pre-trained on VoxCeleb \cite{ravanelli_speechbrain_2021}. This model produces 192-dimensional embeddings that capture speaker-specific acoustic characteristics while being robust to recording conditions and speech content. All audio was resampled to 16 kHz prior to embedding extraction.

We formulated our speaker identification task as a binary classification problem: for each utterance and each enrolled speaker (candidate), predict whether the candidate is the true speaker. This formulation generates multiple candidate-utterance pairs per utterance, with exactly one positive (correct) match per utterance.

\begin{figure}[ht]
\centering
\caption{Structured classroom speaker-identification prompt.}
\label{fig:LLMprompt}
\begin{lstlisting}
system_prompt: |
  Identify speakers from a classroom transcript.

prompt: |
  I have a diarized transcript and I want you to identify who is speaking. 
  The conversation is from a 4th grade math classroom.
  
  Provide a list of dictionaries mapping utterance index to speaker index 
  for any turns you can confidently identify and an explanation of your 
  thought process.

  Only include turns that you can CONFIDENTLY identify as either the teacher 
  or a specific student. Each name should exactly match one of the speaker 
  indexes provided below.
  
  Teacher:
  ${teacher_name}

  Students:
  ${student_names}

  Transcript:
  ${transcript}

parameters:
  - name: teacher_name
    default: |
      - 0: Anne Finch
  - name: student_names
    default: |
      - 1: Anne Sparrow
      - 2: Rob Hummingbird
  - name: transcript
    default: |
      Hi Bob, do you know 3+1?
      Yes Alice, I do it's 4
      Is that right, Mr. Eagle?
\end{lstlisting}
\end{figure}

Further, we engineered a comprehensive feature set capturing multiple aspects of speaker similarity:
\begin{itemize}
    \item \textbf{Utterance-to-Enrollment Similarity}: For each candidate speaker, we computed cosine distances between the utterance embedding and all voice enrollment embeddings for that candidate. We extracted summary statistics including mean, median, 10th percentile, 90th percentile, minimum, and maximum distances. Lower cosine distances indicate higher similarity.
    \item \textbf{Speaker Group Features}: Utterances were initially grouped by their transcript-based TranscribeMe speaker labels (e.g., ``S1'', ``S2'' for students). For each speaker group, we computed:
    \begin{itemize}
        \item Number and proportion of utterances in the group
        \item Inter-utterance similarity within the group (mean and variance of pairwise cosine similarities).
        \item Similarity between each candidate enrollment and the speaker group's mean embedding.
        \item Similarity between each candidate enrollment and the longest utterance in the group.
    \end{itemize}
    \item \textbf{Intra-Group Utterance Distances}: For each utterance, we computed distances to other utterances assigned to the same speaker group, capturing consistency of speaker identity within transcript-labeled clusters.
    \item \textbf{LLM-Inferred Speaker}: We incorporated a binary feature indicating whether GPT-5-mini (using transcript context) inferred the current candidate as the utterance's speaker using the prompt in Figure~\ref{fig:LLMprompt}. For candidates identified by the LLM, we also computed distance statistics to other utterances the LLM assigned to that same speaker across the course. 
    \item \textbf{Utterance Duration}: Duration (in seconds) of each utterance, as speaker identification accuracy tends to increase with longer utterances.
\end{itemize}

All features with missing values (e.g., when LLM inference was unavailable) were filled with a sentinel value of -999 to allow the gradient boosting model to learn separate decision rules for missing data.

\subsection{Model Architecture and Training}

We selected XGBoost, a gradient boosting decision tree algorithm, as our primary classification model due to its strong performance on tabular data with heterogeneous features, natural handling of missing values, and computational efficiency \cite{chen_xgboost_2016}. Given our task formulation, the model predicts a probability that each candidate speaker is the true speaker for a given utterance.

\begin{table}[ht]
\centering
\caption{Hyperparameter Search Space}
\label{tab:hyperparams}
\begin{tabularx}{\columnwidth}{lX}
\toprule
\textbf{Parameter} & \textbf{Range / Scale} \\
\midrule
Number of estimators & 200--1200 (step = 100) \\
Learning rate & 0.01--0.2 (log scale) \\
Maximum depth & 3--10 \\
Minimum child weight & 1.0--10.0 \\
Subsample ratio & 0.6--1.0 \\
Column subsample ratio & 0.6--1.0 \\
Gamma (minimum loss reduction) & 0.0--0.5 \\
L1 regularization ($\alpha$) & $10^{-8}$ to $10^{-1}$ (log scale) \\
L2 regularization ($\lambda$) & $10^{-2}$ to 10.0 (log scale) \\
Scale positive weight & 1.0--100.0 \\
\bottomrule
\end{tabularx}
\end{table}

We performed Bayesian hyperparameter optimization using Optuna with 50 trials \cite{akiba_optuna_2019}. The optimization process used a nested cross-validation strategy:
\begin{enumerate}
    \item \textbf{Outer Loop (Leave-One-Session-Out)}: For each of the eight annotated sessions, we held out that session as a test set and used the remaining seven sessions for training and validation.
    \item \textbf{Inner Loop (Validation Split)}: Within the training set, we further separated one additional session as a validation set for early stopping, ensuring the test session remained completely unseen during hyperparameter selection.
\end{enumerate}

For our training objective, we maximized Top-3 student accuracy (the proportion of student utterances for which the true speaker appeared in the Top-3 predicted candidates) averaged across all folds. The hyperparameter search space is detailed in Table~\ref{tab:hyperparams}. The best hyperparameters identified were: 

\begin{itemize}
    \item n\_estimators = 800
    \item learning\_rate = 0.043
    \item max\_depth = 3
    \item min\_child\_weight = 8.13 
    \item subsample = 0.87 
    \item colsample\_bytree = 0.72
    \item gamma = 0.14
    \item reg\_alpha = 0.086
    \item reg\_lambda = 3.22
    \item scale\_pos\_weight = 41.21
\end{itemize} 

After hyperparameter selection, we trained a final production model on all eight annotated sessions using the optimal hyperparameters. We reserved 20\% of utterances (stratified by session) as an evaluation set for early stopping during this final training phase. Further, XGBoost outputs were calibrated using sigmoid calibration (Platt scaling) fitted on leave-one-session-out predictions \cite{platt_probabilistic_2000}. This ensured predicted probabilities reflected true likelihood of correct speaker assignment, facilitating principled decision-making for speaker assignment thresholds.

\begin{figure}[htbp]
    \centering
    \label{fig:model_diagram}
    \caption{Pipeline for Classroom Session Analysis}
    \includegraphics[width=0.45\textwidth]{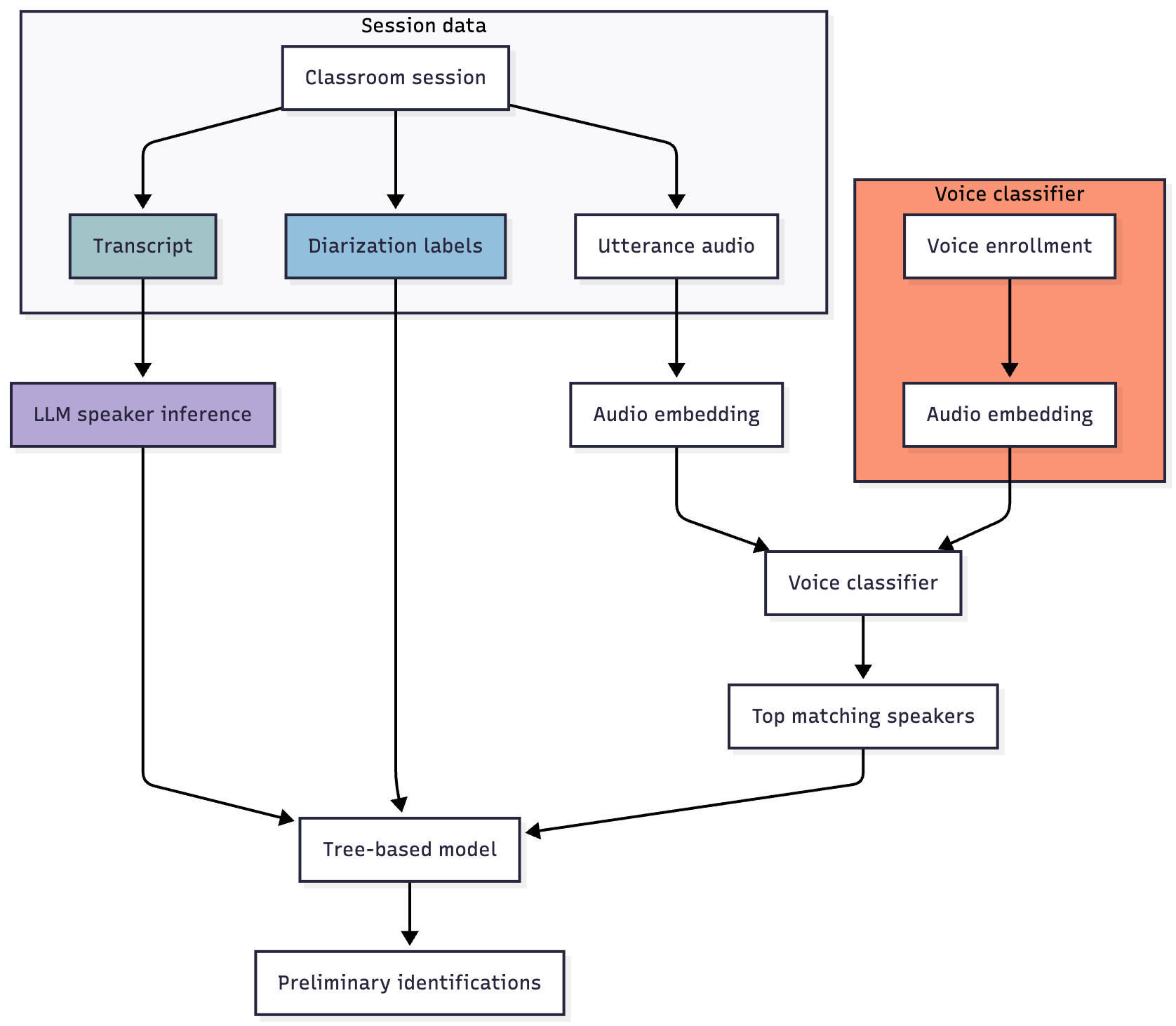}
    {\raggedright\footnotesize System architecture for classroom speaker identification. The figure depicts the pipeline, which integrates transcript inference from LLMs, diarization labels, utterance audio embeddings, and voice classification using similarities to voice enrollments to produce the final speaker probabilities. \par}
\end{figure}

\section*{Results}
\setcounter{subsection}{0}

\subsection{Model Evaluation}
We evaluated model performance using leave-one-session-out cross-validation across all eight annotated sessions. For each session, we trained on the remaining seven sessions and predicted on the held-out session, ensuring completely independent evaluation.

In evaluating model performance, we employed a set of primary metrics designed to capture both overall and role-specific classification accuracy. \textbf{Overall Accuracy} was defined as the proportion of utterances correctly attributed to the true speaker, based on the candidate with the highest predicted probability. To assess role differentiation, \textbf{Student vs. Teacher Accuracy} measured the proportion of utterances correctly classified according to speaker role. Within the subset of student utterances, \textbf{Student Identification Accuracy} quantified the proportion correctly assigned to the specific student, thereby reflecting the model’s ability to distinguish among multiple student speakers. Finally, \textbf{Top-k Accuracy} provided a more flexible measure of predictive reliability by calculating the proportion of utterances in which the true speaker appeared within the top $k$ predicted candidates, with values reported for $k = 1, 3, 5$ and $10$. This suite of metrics offers a comprehensive framework for evaluating both fine-grained speaker identification and broader role-level classification.

\begin{table*}[t] 
    \centering
    \caption{Identification Accuracy}
    \label{tab:identification_accuracy}
    \renewcommand{\arraystretch}{1.2} 
    \begin{tabular}{l ccc c cccc}
        \toprule
        \multirow{2}{*}{\textbf{Utterance duration (s)}} & \multicolumn{3}{c}{\textbf{\# Utterance}} & \multicolumn{1}{c}{\textbf{Student vs Teacher}} & \multicolumn{4}{c}{\textbf{Student vs Student}} \\
        \cmidrule(lr){2-4} \cmidrule(lr){5-5} \cmidrule(lr){6-9}
         & Total & Student & Teacher & Accuracy & Top 1 & Top 3 & Top 5 & Top 10 \\
        \midrule
        (0-1] & 806 & 510 & 296 & 98.0\% & 41.0\% & 63.1\% & 71.8\% & 82.4\% \\
        (1-3] & 831 & 351 & 480 & 99.6\% & 55.0\% & 76.1\% & 82.1\% & 90.6\% \\
        (3-5] & 326 & 90 & 236 & 99.4\% & 62.2\% & 81.1\% & 84.4\% & 90.0\% \\
        (5-10] & 351 & 55 & 296 & 100.0\% & 74.5\% & 90.9\% & 94.5\% & 96.4\% \\
        $>$10 & 487 & 24 & 463 & 100.0\% & 79.2\% & 95.8\% & 95.8\% & 100.0\% \\
        \midrule
        \textbf{Overall} & \textbf{2,801} & \textbf{1,030} & \textbf{1,771} & \textbf{99.3\%} & \textbf{50.3\%} & \textbf{71.4\%} & \textbf{78.2\%} & \textbf{87.0\%} \\
        \bottomrule
    \end{tabular}
    \vspace{2ex}
    \\
    \footnotesize \textit{Note.} Speaker identification accuracy of our XGBoost model at different utterance lengths. Total number of utterances (2,801) reflects a reduction of the total utterances in sample (3999) as we only attempt to identify utterances for which a speaker could be identified.
\end{table*}

We also stratified performance by utterance duration, as acoustic speaker identification typically improves with longer audio samples.

\subsection{Baseline Comparison}

To quantify the performance gain achieved by our multimodal framework, we compared the final XGBoost classifier against a baseline "Acoustic-Only" approach. The baseline method assigned speaker identity solely based on the lowest cosine distance between the utterance embedding and the voice enrollment samples (ECAPA-TDNN pre-trained on VoxCeleb), without the benefit of transcript-based features or metadata-aware re-ranking.

As shown in Table~\ref{tab:performance_comparison}, the multimodal XGBoost classifier yielded meaningful improvements. The most dramatic gains were observed in Student Identification, where the proposed model improved accuracy by 32 percentage points (from ~39\% to ~71\%) compared to the acoustic baseline.

\begin{table}[H]
\centering
\caption{Top-1 Performance Comparison: Acoustic Baseline vs. Multimodal XGBoost}
\label{tab:performance_comparison}
\begin{tabularx}{\columnwidth}{X c c c}
\toprule
\textbf{Metric} & \textbf{\begin{tabular}{@{}c@{}}Acoustic Baseline\\(VoxCeleb Only)\end{tabular}} & \textbf{\begin{tabular}{@{}c@{}}Multimodal\\XGBoost\end{tabular}} & \textbf{Improvement} \\
\midrule
Student~vs.~Teacher & 88.0\% & \textbf{99.3\%} & +11.3 pp \\
Exact Student ID (All) & 39.0\% & \textbf{50.3\%} & +10.7 pp \\
Exact Student ID ($>$5s) & 64.9\% & \textbf{76.9\%} & +12.0 pp \\
\bottomrule
\end{tabularx}
\footnotesize{\emph{Note: pp = percentage points.}}
\end{table}

The model showed similar gains in distinguishing between roles, boosting Student vs. Teacher accuracy from 88.0\% to a near-perfect 99.3\% (+11.3 percentage points). Notably, this performance advantage persisted even in longer utterances ($>$5s), where the acoustic signal is typically more stable. In these cases, the multimodal approach improved identification accuracy from 64.9\% to 76.9\%, a gain of 12.0 percentage points. These results suggest that even when acoustic conditions are favorable, the addition of semantic context and metadata provides a necessary lift to resolve ambiguities in identifying child speakers.

\subsection{Findings}

Our analysis of the $2,801$ utterances reveals that multimodal speaker identification in classrooms performance is heavily dependent on utterance duration and speaker role. As detailed in Table~\ref{tab:identification_accuracy}, the dataset is skewed toward short interactions, with over 58\% of utterances lasting 3 seconds or less. Despite the challenging distribution, the model demonstrates high ability to distinguish distinct speaker roles and to identify speakers in longer turns. 

\textbf{Teacher vs. Student Differentiation} The system achieved near-perfect accuracy in distinguishing between teachers and students, with an overall accuracy of \textbf{99.3\%}. Performance in this category was robust even for the shortest utterances (0-1s), achieving \textbf{98.0\%} accuracy and reached \textbf{100\%} accuracy for all utterances longer than 5 seconds. 

\textbf{Student Identification} Identifying individual students proved more challenging than role detection, particularly for short utterances. The overall Top-1 accuracy for specific student identification was \textbf{50.3\%}. However, this metric is clearly impacted by the brevity of classroom dialogue. 

\begin{itemize}
    \item \textbf{Impact of Duration}: Identification accuracy scales strictly with utterance length. For substantive turns (longer than 10 seconds), the model achieve a \textbf{79.2\%} Top-1 accuracy and a \textbf{95.8\%} Top-3 accuracy. Conversely, short utterances (0-1s) showed a Top-1 accuracy of only \textbf{41.0\%}.
    \item \textbf{Top-k Performance}: The model proves highly effective as a candidate filtering system. While Top-1 accuracy for utterance under 3 seconds hovered between 41-55\%, the Top-5 accuracy for these same segments rose to \textbf{71.8-82.1\%}. Across all durations, the correct speaker was found withing the top 10 candidates 87.0\% of the time, and 100\% of the time for turns over 10 seconds.
\end{itemize}

These results indicate that while short, context-poor utterances remain an adversarial frontier, the multimodal architecture effectively isolates the correct speaker for the substantive contributions necessary for analyzing student reasoning and discourse.

\section*{Discussion}
\setcounter{subsection}{0}

Critically, we find that the model performs best where it matters most for educational feedback. The near-perfect discrimination between teacher and student roles (99.3\%) immediately enables the reliable automation of evaluation of teacher-only metrics, such as ``Uptake'' \cite{demszky_measuring_2021} or ``Questioning'' \cite{demszky_automated_2025}. Furthermore, the model’s performance scales positively with the semantic richness of the turn. For utterances exceeding 10 seconds—segments that typically represent students articulating mathematical reasoning or complex ideas—the model achieves a Top-3 accuracy of 95.8\%. This indicates that while the system may struggle with fleeting backchannels, it successfully recovers the ``high-value'' contributions necessary for tracking individual student mastery and engagement.

\subsection{The Teacher-Student Performance Gap}
Our results indicate a distinct performance gap between teacher and student identification, with teacher identification achieving near-perfect accuracy even at shorter durations. This disparity is likely driven by two convergent factors: data abundance and acoustic model bias.

First, the classroom environment is inherently imbalanced. As shown in Table~\ref{tab:descriptives}, teachers consistently dominate the acoustic space, with average turn lengths (4.88s–18.8s) overwhelmingly exceeding those of students (0.75s–3.11s). This provides the model with substantially more acoustic information for teachers, resulting in more robust centroids for the target speaker.

Second, the acoustic embedding model used (ECAPA-TDNN) was pre-trained on VoxCeleb, a dataset comprised almost exclusively of adult speech. As noted in the introduction, children’s speech exhibits higher spectral variability and higher fundamental frequencies than adult speech. Consequently, the pre-trained feature extractor is likely operating in-domain for teachers (adults) but out-of-domain for students (children), leading to the observed degradation in student-specific accuracy. Our baseline comparison confirms that while acoustic embeddings alone (VoxCeleb) are insufficient for distinguishing between children (39\% accuracy), they provide a necessary foundation that, when combined with semantic features, allows the classifier to reach 71\% accuracy. We attempted to solve for this issue by using versions of the model fine-tuned with children's voice data - following Tabatabaee et. al. \cite{tabatabaee_ft-boosted_2025} - but found that the model performed no better (and in a few instances worse) for these classrooms. 

\subsection{The ``Short Utterance'' Frontier}

The drop in accuracy for utterances under 3 seconds (Top-1 41.0\%–55.0\%) highlights the physical limitations of acoustic biometrics. 
Speaker embeddings require sufficient phonetic variation to form a unique signature. 
Very short student utterances in K-12 classrooms often consist of ``backchanneling'' (e.g., ``Yeah'', ``Okay'') or rapid peer-to-peer overlapping speech. 
Utterance length as well as high environmental noise impact x-vector based speaker identification models, with duration impacting model training more than testing \cite{magrin2024effect, gusev2020deep}. Magrin-Chagnolleau et. al. \cite{magrin2024effect} show that the biggest jump in model performance occurs between an utterance length of 1 and 2 seconds. This is validated in our findings as well, which shows that across all `Top-k' experiments, the largest jump in performance is from an utterance duration of 0-1 second to 1-3 seconds (see Table-\ref{tab:identification_accuracy}). These segments lack the temporal duration required to extract stable x-vectors or ECAPA-TDNN embeddings. 

However, the high Top-5 accuracy (71.8\%–82.1\%) for these short turns suggests that while the model cannot pinpoint the exact student with high confidence, it successfully narrows the search space. This validates the multimodal approach: the acoustic model filters the candidates, potentially allowing the LLM’s semantic context to make the final disambiguation in instances where transcripts capture specific addressee cues (e.g., ``Good job, Michael''). This narrowed search space also vastly reduces the human effort needed to generate high-quality ground truth labels in future work.  

\section*{Conclusion}

This study evaluates a multimodal framework for speaker identification in the ``perfect storm'' of K-12 mathematics classrooms—an environment characterized by non-stationary noise and highly variable child speech. By integrating ECAPA-TDNN acoustic embeddings with LLM-inferred semantic anchors, we demonstrate a robust method for attributing classroom discourse.

Our findings show that this approach solves the critical task of differentiating teacher from student speech with over 99\% accuracy, a foundational requirement for automated instructional feedback systems. While identifying specific students remains challenging in short, sub-second utterances, the model achieves high reliability (95.8\% Top-3 accuracy) for longer, substantively rich turns. This suggests that the ``linguistic signature'' provided by transcript context significantly complements acoustic biometrics when sufficient semantic information is present.

These results advance the feasibility of scalable, equitable classroom analytics. By accurately tracking ``who said what'', such systems can move beyond aggregate class metrics to provide granular insights into individual student participation and teacher-student interaction patterns. Future work should focus on improving resolution for short utterances to capture the full spectrum of rapid-fire classroom discourse.

\section*{Acknowledgment}
We thank the Gates Foundation, the Walton Family Foundation, and the Chan Zuckerberg Initiative for their generous financial support.

We also gratefully acknowledge the contributions of Talya Lebson, who provided careful and consistent annotation of the speaker identification dataset. Their diligence in applying the coding framework and attention to detail were essential to ensuring the reliability and validity of the data used in this study. 

\bibliographystyle{IEEEtran}
\bibliography{references}

\end{document}